\documentstyle[aps,prl,twocolumn,psfig,floats,url]{revtex}
\begin{document}
\preprint{\vbox{\hbox{UCB-PTH-01/31}},
  \vbox{LBNL-48787}}
\draft
\wideabs{
\title{Not Even Decoupling Can Save Minimal Supersymmetric SU(5)}
\author{Hitoshi Murayama and Aaron Pierce}
\address{
Department of Physics, 
University of California, 
Berkeley, CA~~94720, USA;\\ 
Theory Group, 
Lawrence Berkeley National Laboratory, 
Berkeley, CA~~94720, USA}

\date{\today}
\maketitle


\setcounter{footnote}{0}
\setcounter{page}{1}
\setcounter{section}{0}
\setcounter{subsection}{0}
\setcounter{subsubsection}{0}

\begin{abstract}
  We make explicit the statement that Minimal Supersymmetric SU(5) has
  been excluded by the Super-Kamiokande search for the process $p \to
  K^{+} \overline{\nu}$.  This exclusion is made by first placing
  limits on the colored Higgs triplet mass, by forcing the gauge
  couplings to unify.  We also show that taking the superpartners of the
  first two generations to be very heavy in order to avoid flavor
  changing neutral currents, the so-called ``decoupling'' idea, is
  insufficient to resurrect the Minimal SUSY SU(5).  We comment on
  various mechanisms to further suppress proton decay in SUSY SU(5). Finally, we address the contributions to proton decay from gauge boson
exchange in the Minimal SUSY SU(5) and flipped SU(5) models.
\end{abstract}
}
\vskip 0.3in

\section{Introduction}\label{sec:intro}

Proton decay would be a smoking gun signature for Grand Unified
Theories (GUTs).  Unfortunately, no such signal has been seen.  In fact, very
strong experimental limits have been set for this process, placing the
minimal GUTs in a very precarious position.  Super
Kamiokande has set a lower limit on the proton lifetime in the channel
$p \rightarrow K^{+} \overline{\nu}$ of $6.7 \times 10^{32}$ years at
the 90\% confidence level \cite{SKPDecay}.  This has already placed
stringent constraints on SU(5). We explicitly review the situation for
proton decay in minimal supersymmetric SU(5) and show that the theory
is easily excluded.
   
Because the minimal case is so easily excluded, one might attempt to
tweak the parameters of the theory in some way to push the proton
lifetime upwards.  One such proposed adjustment can be motivated by
the supersymmetric (SUSY) flavor problem.  The numerous parameters of the soft SUSY-breaking sector are {\it a priori} arbitrary, and generically the SUSY-breaking sector will give rise to phenomenologically dangerous flavor-changing
neutral current effects.  One proposal for avoiding such neutral
current difficulties is to decouple the first two generations of
superpartners by making them very heavy \cite{Cohen,Feng,HKN}.  The lore
has been that such a decoupling would also push predictions for proton
decay to an acceptable level.  We show that this is not the case, and
such a modification of the parameters of supersymmetric SU(5) is not
enough to save it.  After painting this bleak picture for the minimal SU(5) theory, we review variations on the theory that are not yet excluded.  Finally, we study the issue of the contributions to proton decay from $X$ and $Y$ gauge boson exchange.

\section{Dimension Five Decay Mechanism}\label{sec:decay}

The $p \rightarrow K^{+} \overline{\nu}$ channel is predicted to be
dominant for supersymmetric SU(5) theories
\cite{Ellis,GotoNihei,Nath1,Nath2,Hisano}. We concentrate on this
channel here.  This channel is enough to exclude the minimal SUSY
SU(5).

The $p \rightarrow K^{+} \overline{\nu}$ decay results from dimension
5 operators, and the associated dressing diagram\cite{WeinbergSakai},
shown in Fig. 1. The dimension five operators come from colored Higgs
triplet exchange, and arise from the following terms in the
superpotential:
\begin{eqnarray}
\lefteqn{W_{Y}=h^{i}Q_{i}u^{c}_{i}H_{f} +
V_{ij}^{*}f^{j}Q_{i}d_{j}^{c}\overline{H}_{f} + f_{i} e_{i}^{c} L_{i}
\overline {H}_{f}} \nonumber \\  
& &+ \frac{1}{2} h^{i} e^{i\phi_{i}} Q_{i} Q_{i} H_{C} +
V_{ij}^{*}f^{j} Q_{i} L_{j} \overline{H}_{C}+ h^{i} V_{ij} u_{i}^{c}
e^{c}_{j} H_{C} \nonumber \\ 
& &+ e^{-i \phi_{i}} V_{ij}^{*} f^{j} u_{i}^{c} d_{j}^{c}
\overline{H}_{C}. 
\label{superpot}
\end{eqnarray}
Here, the $H$ and the $\overline{H}$ represent the two different Higgs
multiplets that give the up and down type quarks their masses.  The
$H_{f}$ is the doublet, while the $H_{C}$ is the colored Higgs
triplet.  All fields are superfields.  $h^i$ and $f^{j}$ are Yukawa
couplings, $V_{ij}$ is a CKM matrix element, and $\phi_{i}$ is a
phase, which is subject to the constraint $\phi_{1} +\phi_{2}+
\phi_{3}=0$.  We will address the decays that result from Higgs
triplet exchange in some detail in the following sections.

\begin{figure}[b]
\centerline{\psfig{file=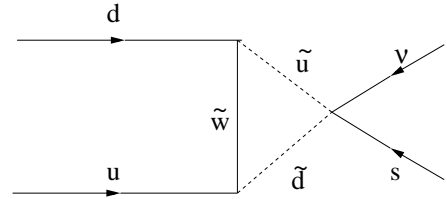,height=1in}}
\vspace{.2in}
\label{fig:dim5}
\caption{The dimension five operator results from the exchange of the colored Higgs triplet.  The super-particles are then removed from the initial state by chargino exchange.  Wino exchange is shown here, but there is an analogous diagram which involves higgsino exchange.}
\end{figure}
  
\section{RGE Arguments}\label{sec:RGE}
In a grand unified theory, we expect that the gauge couplings should
precisely unify.  Particles near the GUT scale provide
corrections to the renormalization group trajectories of the coupling
constants.  These corrections are calculable in terms of the quantum numbers and the masses of the GUT scale particles.  Therefore, by imposing the constraint that the gauge couplings
exactly unify, we can make statements about the high-energy structure
of the theory.  This technique has already appeared in the literature
\cite{couplings,Hisano,Hisano2}.  However, these papers were written
when the knowledge of the strong coupling, $\alpha_s$, was less precise.
Measurements at LEP and SLD have allowed a substantially more precise
determination of $\alpha_s(m_{Z})$. Utilizing this knowledge, we can
dramatically improve the constraint on the mass of the colored Higgs
triplet, $M_{H_C}$.  Constraining the Higgs triplet mass is of
particular importance since it mediates the dominant decay of the
proton.

The colored Higgs triplets are not the only new particles at the GUT
scale.  We expect to have a $\Sigma_{24}$ Higgs, new vector bosons
(denoted collectively by $V$), in addition to the colored Higgs
triplet, $H_{C}$, near the GUT scale.  One might think that it would
be impossible to determine $M_{C}$ without knowledge of $M_{\Sigma}$,
and $M_{V}$.  However, by examining the RGEs for the gauge couplings
at one loop (neglecting the Yukawa couplings):

\begin{eqnarray}
\lefteqn{
\alpha_{3}^{-1}(m_Z)=\alpha_5^{-1}(\Lambda) + \frac{1}{2 \pi} \Bigl[
(-2 -\frac{2}{3}N_{g}) \log \frac{m_{SUSY}}{m_Z} } \nonumber  \\ 
& &~+ (-9+2N_{g})\log \frac{\Lambda}{m_{Z}} -4 \log
\frac{\Lambda}{M_{V}}  \nonumber \\ 
& &~+3 \log \frac{ \Lambda}{M_{\Sigma}} + \log
\frac{\Lambda}{M_{H_C}}\Big],\nonumber \\ 
\lefteqn{
\alpha_{2}^{-1}(m_Z)= \alpha_5^{-1}(\Lambda) + \frac{1}{2 \pi} \Bigl[
(-\frac{13}{6} -\frac{2}{3}N_{g}) \log \frac{m_{SUSY}}{m_Z} }\nonumber
\\ 
& &~+ (-5+2N_{g})\log \frac{\Lambda}{m_{Z}} -6 \log
\frac{\Lambda}{M_{V}} + 2\log \frac{ \Lambda}{M_{\Sigma}} \Bigr],
\nonumber \\ 
\lefteqn{
\alpha_{1}^{-1}(m_Z)= \alpha_5^{-1}(\Lambda) + \frac{1}{2 \pi} \Bigl[
(-\frac{2}{3}N_{g}-\frac{1}{2}) \log \frac{m_{SUSY}}{m_Z} } \nonumber \\ 
& &~+(\frac{3}{5}+2N_{g})\log \frac{\Lambda}{m_{Z}} -10 \log
\frac{\Lambda}{M_{V}} + \frac{2}{5} \log
\frac{\Lambda}{M_{H_C}}\Bigr],
\end{eqnarray}
we find that we can eliminate $M_{\Sigma}$ and $M_{V}$ by taking a
judicious combination of the couplings \cite{couplings}.  In the case
of the above RGEs, neglecting the Yukawa couplings, we find:
\begin{eqnarray}
\label{eqn:HiggsRGEMass}
3\alpha_{2}^{-1}(m_{Z})-2\alpha_{3}^{-1}(m_{Z})-\alpha_{1}^{-1}({m_{Z}})= \nonumber \\
\frac{1}{2 \pi}
\Bigl(\frac{12}{5} \log \frac{M_{H_C}}{m_{Z}} - 2 \log
\frac{m_{SUSY}}{m_{Z}}\Bigr). 
\end{eqnarray}
We can invert the above equation to determine the colored Higgs mass
independently of the other masses at the GUT scale.

This one loop example gives the basic procedure.  In the numerical
calculation that follows, we use the two loop RGEs for the gauge and Yukawa
couplings between the SUSY scale and the GUT scale, which can be
found, for example, in \cite{Einhorn}.  Here, the SUSY scale is defined as the mass scale above which all superpartners contribute to the RGEs.  We include only the Yukawa couplings of the third generation, all others are neglected.  We use
one loop RGEs for all running between $m_{Z}$ and the SUSY scale.  We
also include the one loop finite effects at the wino and gluino
threshold, using the results of \cite{Yamada}.  There is no simple
analytic solution for the colored Higgs mass, so we must do a
numerical analysis.

It is further necessary to take into account the splitting of the
supersymmetric particle spectrum.  We make the approximation that all
the supersymmetric particles, aside from the gauginos, are degenerate
at a TeV.  As long as the splitting between the sparticles within each SU(5) multiplet is not too large,
this is a reasonable approximation.  Because the proton decay constraint ends up requiring
scalars to be somewhat heavy, the expected splittings within each
SU(5) multiplet due to the gaugino contribution in the RGE is small.  

From the ratio
between the couplings near the SUSY scale, we expect
$\frac{M_{3}}{M_{2}}$ to be 3.5.  With this approximation, we are
left with $M_{2}$ and tan $\beta$ as free parameters.  In the limits
quoted below, we set $M_{2}$=200 GeV.  We scan over $\tan \beta$
between 1.8 and 4.  Large values of $\tan \beta$ are very bad for
proton decay, and the top Yukawa becomes non-perturbative below 1.8.
In fact, recent results from Higgs searches at LEP \cite{LEPHiggs}
suggest that $\tan \beta > 2.4$.  However, these bounds can probably be avoided by modifying the Higgs sector.\footnote{For example,
  by including a singlet field as in the NMSSM, one can weaken these
  bounds using the larger Higgs self-coupling and/or the invisible
  decay of Higgs into singlet scalars.  The tadpole problem in the
  NMSSM can be avoided even with GUT if the supersymmetry breaking
  originates in gauge mediation at low energies.}  Therefore, we
conservatively scan the interval between 1.8 and 4, a scan between 2.4
and 4 would only make things worse for SU(5).
 
We use the following precision measurements as inputs \cite{PDG}:
\begin{eqnarray}
\alpha_{s_{\overline{MS}}} (m_{Z})&=&.1185 \pm .002 \label{eqn:precisionEM} \\
\sin^{2} \theta_{w_{\overline{MS}}} (m_{Z})&=& .23117 \pm .00016
\label{eqn:precisionW}\\ 
\alpha_{em_{\overline{MS}}}(m_{Z})&=&\frac{1}{127.943 \pm .027}
\label{eqn:precisionS} . 
\end{eqnarray}

All these quantities are given in the $\overline{MS}$ scheme.
However, the step function approximation at particle thresholds is
good only in the $\overline{DR}$ scheme\cite{DRThresh}.  Yukawa
couplings and gauge couplings must therefore be converted from
$\overline{MS}$; the dictionary for this conversion may by found in
reference \cite{DR}.

Operationally, we use a given colored Higgs mass along with the
renormalization group equations to predict the data of
Eqns.~(\ref{eqn:precisionEM},\ref{eqn:precisionW},\ref{eqn:precisionS}). 
We find that SU(5) prediction of exact unification agrees with the data (using a $\chi^{2}$ fit for the one degree of freedom: $M_{H_{C}}$) only for colored Higgs masses of:
\begin{eqnarray}
3.5 \times 10^{14} \leq M_{H_C} \leq 3.6 \times 10^{15} \mbox{GeV}\nonumber \\ 
(90\% \mbox{ confidence level}).
\label{eqn:RGELimit}
\end{eqnarray}
We find that varying $M_{2}$ within a reasonable range (100-400 GeV)
causes a change in the $M_{H_C}$ bounds on the order of 10\%.  The previous upper limit of reference \cite{Hisano2}, was $M_{H_{C}} < 2.4\times 10^{16}$ GeV.  The improvement is largely due to the improvement in the precision on $\alpha_{s}$.  

Note that the
above limit will not be drastically affected in the case where we take
the scalars of the first and second generations to have masses on the
order of 10 TeV.  This is because changing the energy scale of an
entire SU(5) multiplet does not change the unification condition, and hence the RGE bound, at one-loop.  A small sparticle splitting within a multiplet relative of the sparticles masses is especially well motivated if the first and second generation scalars are pushed up to 10~TeV, otherwise a problematic Fayet--Illiopoulos $D$-term \cite{Cohen} is induced. 
This fact will be of use when we move on to discuss the decoupling
scenario in section \ref{sec:decoupling}.

We also note that it is possible to place a constraint on the
combination $(M_{\Sigma} M_{V}^{2})^{1/3}$.  This is done by looking
at the combination $5\alpha_{1}^{-1}-3\alpha_{2}^{-1}-2\alpha_{3}^{-1}$
\cite{couplings}.  We find that this scale is very tightly
constrained:
\begin{eqnarray}
1.7 \times 10^{16} \leq (M_{\Sigma} M_{V}^{2})^{1/3}\leq 2.0 \times
10^{16} \mbox{ GeV}\nonumber \\ 
(90\% \mbox{ confidence level}).
\label{eqn:GUTLimit}
\end{eqnarray}
In what follows, we refer to the scale $(M_{\Sigma} M_{V}^{2})^{1/3}$ as $M_{GUT}$.  Incidentally, the above bounds of Eqns.~(\ref{eqn:RGELimit},\ref{eqn:GUTLimit}), are not uncorrelated. We show the allowed region in the $M_{H_{C}}-M_{GUT}$ plane in Fig. 2.  The bounds that result from projecting the ellipse in the figure on to one of the axes are weaker than those in Eqns.~(\ref{eqn:RGELimit},\ref{eqn:GUTLimit}).  This is because the ellipse is found by performing a fit using a $\chi^{2}$-distribution with two degrees of freedom, whereas the bounds in the equations are found using a $\chi^{2}$-distribution with one degree of freedom.
\begin{figure}
  \centerline{
  \psfig{file=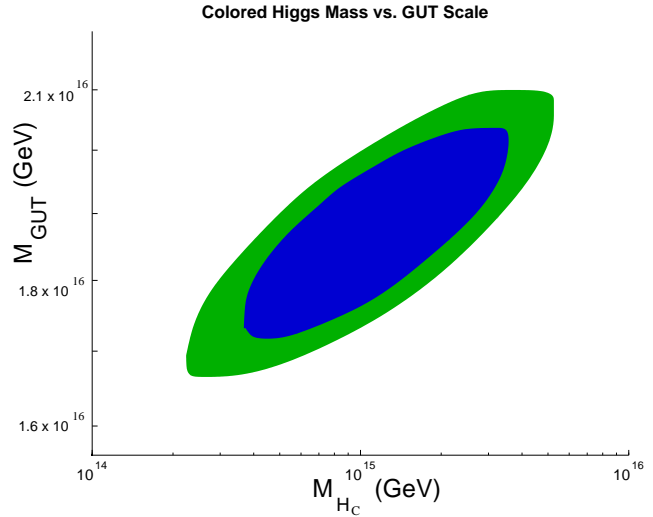,height=3.0in}
  }
\label{fig:corrmat}
\caption{Plot showing 68\% and 90\% contours allowed by the renormalization group analysis for the color Higgs triplet mass, $M_{H_{C}}$, and the GUT scale, $M_{GUT} \equiv (M_{\Sigma} M_{V}^{2})^{1/3}$. }
\end{figure}  

What are the consequences of such a strong limit on the colored Higgs
mass for minimal SUSY SU(5)?  They are not good.  Our calculation of
the proton lifetime follows the methods of reference
\cite{GotoNihei}.\footnote{Our calculation shows an approximate factor
  of two discrepancy with that reference.  Our predicted lifetime is
  shorter, but in any case, it will not affect the qualitative nature
  of our results in any way.}  Although values of $\mu$ on the order
of 800 GeV are favored by the electroweak symmetry breaking condition,
we take $\mu$ as a free parameter in our phenomenological analysis.
We keep the $M_2$ as a free parameter, and determine the other gaugino
masses through the unification condition.  For the scalars, we take
the stop soft masses to be 400 and 800 GeV at the weak scale, and set
the masses of all other SUSY particles to have masses of 1 TeV.  We
neglect squark and slepton mixing, except for the stops.  With these
assumptions in place, we scan over the parameters $\mu$, $M_2$, $\tan
\beta$, and the independent phases $\phi_{1}$ and $\phi_{2}$, to
maximize the lifetime as a function of $M_{H_C}.$ We allow $\tan
\beta$ to vary in the interval $\tan \beta \in (1.8, 4)$; $M_2$ to
vary in the interval $M_2 \in (100,400)$, and $\mu \in (100,1000)$.
We eliminate those points which have a too-light chargino mass, using
the constraint from LEP II \cite{LEPChargino}, $m_{\chi^{+}} > 103.5$
GeV.  The Yukawa couplings are extracted from the central values of
the quark masses listed in reference \cite{PDG}.

In our calculation, we take into account both short and long range
renormalization effects. Yukawa couplings must be run up to the GUT
scale.  The Wilson coefficients of the effective dimension five
operators must be run back down to the SUSY scale.  We use the RGEs
from the appendix of reference \cite{GotoNihei}, ignoring all Yukawa
couplings except for that of the top quark.  The one-loop
renormalization of the Wilson coefficients of the dimension six
operators from the weak scale to 1 GeV can be extracted from reference
\cite{Ellis}.  The renormalization of the Yukawa couplings (quark
masses) from 2 GeV to the weak scale is done to three loops.

Using the newer limit from Super Kamiokande of $6.7 \times 10^{32}$
years (90 \% confidence level), we find that search for proton decay
imposes the constraint:
\begin{equation}
M_{H_C} \geq 7.6 \times 10^{16} \mbox{ GeV}.
\end{equation}
Comparing this equation with Eqn.~(\ref{eqn:RGELimit}), we find that the
minimal SUSY SU(5) theory is excluded by a lot.

It should be noted that this is a very conservative value.  In
particular, this calculation utilizes the traditionally most
conservative value of the hadronic parameter $\beta_{H}= \langle 0 |
u_L u_L d_L | p \rangle = 0.003$ GeV$^{3}$.  Recently, however, there
has been progress on the evaluation of this parameter by the JLQCD
group \cite{JLQCD}.  They find a value, $\beta_{H}=.014 \pm .001$
GeV$^{3}$.  However, this result is to be evaluated at a scale of 2.3
GeV, whereas the value $\beta_{H}= 0.003$ GeV$^{3}$, was to be
utilized at a scale of 1 GeV.  This difference causes the enhancement
of the decay rate to be somewhat less than the naive factor of twenty.
Repeating the above analysis, utilizing the central JLQCD value for
$\beta_{H}$, we find the even more stringent constraint
\begin{equation} 
M_{H_C} \geq 2.0 \times 10^{17} \mbox{ GeV}. 
\end{equation}
This result is in even sharper conflict with Eqn.~(\ref{eqn:RGELimit}).

\section{The Failure of Decoupling}\label{sec:decoupling}

Previous calculations of the proton lifetime have assumed nearly
degenerate scalars at the weak scale, or order 1 TeV in mass.  We made
this same assumption in our calculation in the previous section.  It
seems that one possible escape for the SUSY SU(5) theory with the
minimal field content would be the interesting possibility raised by
reference \cite{Cohen}.  This scenario allows the first and second
generations of scalars to be heavy without severe fine-tuning because they do not affect the
Higgs boson self-energy at the one-loop level.  Even though there is a
naturalness problem at the two-loop level \cite{AM}, the scenario in
\cite{HKN} achieves it without compromising naturalness (the model in
\cite{Feng} does not seem to allow a large splitting).  Since the
proton decay amplitude goes like $m_{\chi}/m_{\tilde{q}^{2}}$, it
seems like we might get a large suppression by making the squarks
ultra-heavy.  However, we will see that even this will not save us.
This point is made clear by looking at the main contributions to
proton decay.  We can write the contributions to $\Gamma(p \rightarrow
K^{+} \overline{\nu})$ as:
\begin{eqnarray}
\label{contributions}
A(p \rightarrow K^{+} \overline{\nu_{e}}) &\approx& [e^{i\phi_{2}}
A_e(\tilde{c}_{L}) + e^{i\phi_{3}} A_{e}(\tilde{t}_{L})]_{LLLL} \nonumber
\\ 
A(p \rightarrow K^{+} \overline{\nu_{\mu}}) &\approx& [e^{\phi_{2}}
A_{\mu}(\tilde{c}_{L}) + e^{i\phi_{3}} A_{\mu}(\tilde{t}_{L})]_{LLLL}
\nonumber \\ 
A(p \rightarrow K^{+} \overline{\nu_{\tau}}) &\approx& [e^{i\phi_{2}}
A_{\tau}(\tilde{c}_{L}) + e^{i\phi_{3}} A_{\tau}(\tilde{t}_{L})]_{LLLL}
\nonumber  \\  
& & +e^{i\phi_{1}}A_{\tau}(\tilde{t}_{R})_{RRRR}. 
\end{eqnarray}
Here, the LLLL subscript refers to the contribution that arises from
dressing the dimension five operator with four left-handed particles,
while RRRR refers to the contribution that arises from dressing the
dimension five operator with for right-handed particles. The RRRR
operator will obviously only have a higgsino piece, and not a wino
piece.  As such, it will only contribute for the $\nu_{\tau}$ case,
where third generation Yukawa couplings allow it to become big
\cite{GotoNihei}.  This contribution was overlooked in earlier
analyses, presumably because the large Yukawa coupling of the top
quark was unanticipated.

When we write the contributions to proton decay as above, it becomes
clear why the decoupling of the first two generations does not save
us.  Although we are able to eliminate the contribution due to the
exchange of the $\tilde{c}$ squark, the contribution due to the stop
still persists.  In the limit of the very heavy scharm, we can rewrite
Eqn.~(\ref{contributions}) as:
\begin{eqnarray} 
\label{3contributions}
A(p \rightarrow K^{+} \overline{\nu_{e}}) &\approx& e^{i\phi_{3}}
A_{e}(\tilde{t}_{L})_{LLLL} \nonumber \\ 
A(p \rightarrow K^{+} \overline{\nu_{\mu}}) &\approx& e^{i\phi_{3}}
A_{\mu}(\tilde{t}_{L})_{LLLL}  \nonumber \\ 
A(p \rightarrow K^{+} \overline{\nu_{\tau}}) &\approx& e^{i\phi_{3}}
A_{\tau}(\tilde{t}_{L})_{LLLL}
+e^{i\phi_{1}}A_{\tau}(\tilde{t}_{R})_{RRRR}.  
\end{eqnarray}

We have not helped matters by making the scharm heavy.  In fact, we
are in many ways worse off, because we cannot use the $\tilde{c}$
contribution to help cancel off the large RRRR contributions to $p
\rightarrow K^{+} \overline{\nu}_{\tau}$.  The basic point is that
proton decay has an important contribution from the exchange of third
generation sparticles.  This causes the decoupling idea to fail.  We
present our quantitative results below.

We took the third generation sparticles to weigh 1 TeV at the weak
scale, except for the top squarks, which, as before, we give soft
masses of 800 and 400 GeV at the weak scale.  We take the first two
generation sparticles at 10 TeV.  In the case that the squarks and
sleptons are much heavier than the chargino, the triangle loop gives a
contribution that goes like ${m_{\chi}}/{m_{\tilde{q}}^{2}}$.
Therefore, placing them at 10 TeV effectively decouples them, by
suppressing their contribution to the amplitude by a factor of $10^2$.

Again, we scan over the relevant parameter space to determine the
maximum proton lifetime. However, there are fewer free parameters than
the case where all generations of sparticles contribute.  In
particular, we can already see that the phase $e^{i\phi_{23}}=
e^{i\phi_{2}}/e^{i\phi_{3}}$ drops out completely.  What is more, if we
wish to conservatively maximize the lifetime predicted by such a
theory, we find that $\phi_{13}$ is determined to be $\pi$.  This
effects the largest possible cancellation between the two
contributions to $A(p \rightarrow K^{+} \overline{\nu_{\tau}})$.  The
remaining free parameters in our calculation are $ \tan \beta$,
$M_{2}$, and $\mu$.  Because the RRRR contribution that arises from
higgsino exchange is much larger than the contribution from wino
exchange, it turns out the the amplitude does not depend strongly on
the value of $M_{2}$. When the decay rate is higgsino-exchange
dominated, nearly the entire branching ratio is to $K^{+}
\overline{\nu}_{\tau}$. We plot the proton lifetime in the $M_2$-$\mu$
plane in Fig. 3 for a fixed value of $\tan \beta$.  There is a
relatively strong dependence on $\tan{\beta}$.  It has long been known
that the large $\tan \beta$ region is bad for proton decay.  This can
be seen explicitly in Fig. 4, where  we show the region between $\tan
\beta$ of 1.8 and 20.  

\begin{figure}
  \centerline{
  \psfig{file=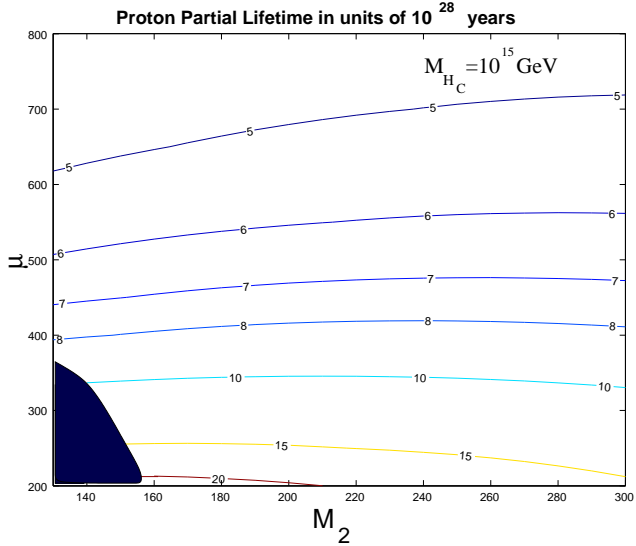,height=3.2in}
  }
\label{fig:muM2}
\caption{A contour plot of the proton partial lifetime, $\tau(p \to K^{+} \bar{  \nu})$, in the case where the
  $1^{st}$ and $2^{nd}$ generation scalars are taken to be 10 TeV.
  The third generation scalars are taken to have masses order 1 TeV,
  except for the stops, which are given soft masses of 800 GeV and 400
  GeV.  We fix tan $\beta$ to be 2.1.  Note that the lifetime is
  approximately proportional to $\mu$, and essentially independent of
  $M_{2}$.  The shaded region is excluded by chargino searches at LEP
  II.  Lifetimes for other values of $M_{H_{C}}$ can be found by noting that the lifetime goes as $M_{H_{C}}^{2}$.}
\end{figure}

\begin{figure}
  \centerline{
  \psfig{file=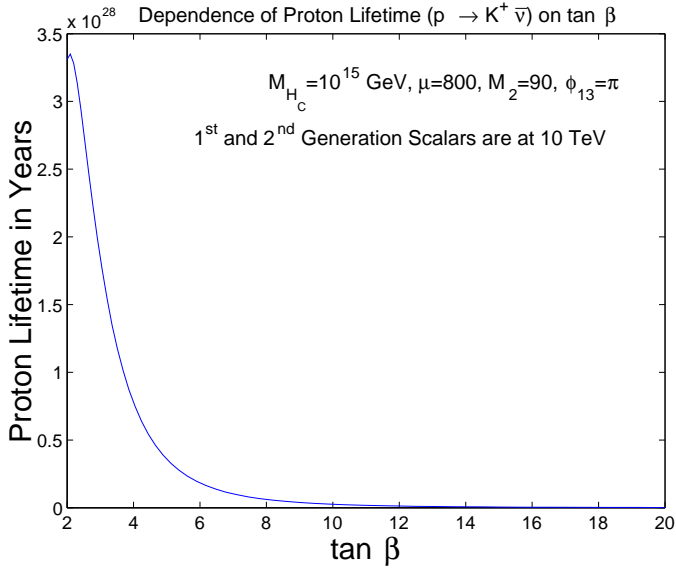,height=3.2in}
  }
\label{fig:tanbeta}
\caption{A plot of proton partial lifetime, $\tau(p \to K^{+} \bar{\nu})$, vs. $\tan \beta$.  Top squark masses are 400 GeV and 800 GeV, while all other $3^{rd}$ generation sparticles have masses are set to 1 TeV.  All other
  variables are fixed as stated.  It is seen that the lifetime peaks
  for values of tan $\beta$ slightly greater than 2 in this case.  Lifetimes for other values of $M_{H_{C}}$ can be found by noting that the lifetime goes as $M_{H_{C}}^{2}$.} 
\end{figure}

The maximum value of the proton lifetime was found by scanning the
parameter space from $\mu \in (80,400)$, $M_2 \in (100,400)$, $\tan
\beta \in (1.8, 3.0)$.  As before, we eliminate those points which
have a too-light chargino mass, using the constraint from LEP II
\cite{LEPChargino}, $m_{\chi^{+}} > 103.5$ GeV.  Using the maximum
value of the colored Higgs mass allowed by our RGE analysis (at 90\%
confidence), $3.6 \times 10^{15}$ GeV,
we find that the maximum value
of the proton partial lifetime is:
\begin{equation}
\tau(p \rightarrow K^{+} \overline{\nu}) \leq 2.9 \times 10^{30} \mbox{ yrs}.  
\end{equation}
Therefore, even the situation with very heavy first and second
generation scalars is {\it easily} excluded at the 90\% confidence
level.  We should reiterate that our RGE analysis is largely unaffected
by our decoupling the first two generations of particles.  First of
all, we are only separating the sparticles from the third generation
by one decade in energy.  Moreover, we have argued that the splitting within the second generation of superpartners is small, and decoupling entire
generations of superpartners has no effect on the unification condition at one loop.   For
the sake of completeness, we also quote the bound on $M_{H_{C}}$, independent of
the RGE analysis.  We find
\begin{equation} \label{eqn:3genmass}
M_{H_{C}} > 5.7 \times 10^{16} \mbox{ GeV}.
\end{equation} 
The statement that this theory is excluded is equivalent to the statement that the above equation is in conflict with \ref{eqn:RGELimit}.
Again, upon utilization of the JLQCD central value for $\beta_{H}=0.014$ GeV$^{3}$, we find that the maximum proton lifetime is even smaller.  In particular, we find that:
\begin{equation}
\tau(p \rightarrow K^{+} \overline{\nu}) \leq 2.5 \times 10^{29} \mbox{ yrs.},
\end{equation}
making the situation even worse.

\section{Avoiding the Constraint}

We wish to stress that, while things look grim for the minimal SU(5)
theory, our result does not mean that no SU(5) theory is viable.
There exists a host of ideas that allow one to evade the difficulties
outlined in the previous two sections.  They fall into two main
categories.  The first category consists of ideas to evade the
constraints from the RGE arguments.  The second strategy is to somehow
suppress the contribution from the dimension five operators.

In the first strategy, the goal is to push the mass of the colored
Higgs triplet very heavy, thereby suppressing the dimension five
operators.  Then a way must be found to avoid the RGE constraint of
section \ref{sec:RGE}.  To do this, one must include fields that make
additional contributions to the GUT-scale threshold corrections.
Although there are several ways to accomplish this feat, perhaps the
simplest way to do this is to include a second pair of Higgs bosons in
the ${\bf 5}+\overline{\bf 5}$ representation without any Yukawa
coupling to matter multiplets.  However, in this pair one makes the
triplet {\it lighter} than the doublet.  As such, the threshold
corrections to unification from this pair will work in a way opposite
from the correction from the usual Higgs multiplet, and can allow the
original Higgs triplet to be heavier.

The second strategy is to suppress the dimension five operators in
some way.  A number of ideas exist in the literature for accomplishing
this goal.  Most recently, some interesting ways of eliminating the
dimension five operators entirely in an extra-dimensional framework
\cite{ExtraDim} have appeared. Another attempt utilizes a somewhat
complicated Higgs sector, but succeeds in suppressing dimension five
operators or even removing them entirely \cite{BabuBarr}. In general, the dimension-five operators are
sensitive to the mechanism of doublet-triplet splitting, arguably the
least pleasant aspect of GUT.  In some models that achieve the
doublet-triplet splitting in a natural way, dimension-five operators
are eliminated, such as in flipped SU(5)\cite{flipped}.  Yet another method for suppressing the dimension five operators is to somehow suppress the Yukawa couplings between the standard model
fermions and the colored Higgs triplet.  In the past, this might have
been considered the favored mechanism for suppressing proton decay,
simply because there were already problems in the minimal SU(5) with
GUT relationships like $m_{\mu}=m_{s}$.  It was assumed that attempts
to remedy these fermion mass relationships would somehow also remedy
the proton decay problem.  However, since it is now recognized that
there is a dominant contribution from the RRRR operator, which is
proportional to the $3^{rd}$ generation Yukawas, one would have to
modify the flavor structure of the third generation in some way as
well, which is less likely.

Finally, methods exist to suppress the dimension five operators where the two strategies mentioned above are combined.   For example, one mechanism includes an
additional pair of Higgs triplets, $H_{C}'$ and $\overline{H}_{C}'$,
that exist solely to give the original pair of Higgs triplets a mass.
In this case, the operator that arises from integrating out the
$M_{H_{C}} \overline{H_{C}} H_{C}$ term can be forbidden by a
Peccei--Quinn symmetry \cite{WeinbergSakai}.  However, the symmetry
needs to be eventually broken, and it turns out that the RGE bound
constrains the combination relevant for the dimension five operator
\cite{PQsuppression}.  So, something must be added to the model to help avoid this bound.  Inspiration comes from the missing partner model \cite{missingpartner}, which
utilizes a SU(5)-Higgs in the {\bf 75} representation.  This generates an
additional threshold correction that pushes the RGE limit on the
color-triplet Higgs higher \cite{HagiwaraYamada}.  However, the simplest incarnation of the missing partner model model has the problem that the gauge coupling becomes non-perturbative soon above the GUT-scale.  The answer comes in combining the two models: adding the the Peccei--Quinn symmetry to the missing partner model can be used to postpone the peturbativity problem.  The resulting suppression from the symmetry is sufficient to make the dangerous proton decay of the previous sections benign \cite{Hisano,Hisano3}. 

SO(10) models, having more multiplets at the GUT-scale, allow larger
threshold corrections and hence can loosen the bound on the
color-triplet Higgs mass if the threshold correction comes with the
correct sign.  Moreover, there are many color-triplet Higgses which
mix with each other.  Even though suppressing proton decay and
achieving the correct threshold correction often have tension, one can
build models to achieve an overall suppression \cite{BPW}.

\section{Dimension Six Proton Decay}
In general, the dimension six operator arising from X and Y gauge boson exchange provides a less model-dependent decay rate.\footnote{If SU(5) is broken on an orbifold by a boundary
  condition, and if matter fields live on the fixed point where $X$,
  $Y$ bosons vanish, dimension-six operators can be eliminated.  This
  may be viewed as a partial explicit breaking of SU(5) \cite{HMN}.}
With the old evaluation of the hadronic matrix elements, it was
thought that the dimension six operators would be completely out of
reach for the foreseeable future.  However, with the updated value of
the hadronic matrix element from the JLQCD collaboration, the
prospects of detection are slightly less bleak.  Reference
\cite{hisanorecent} has already re-examined this question for the
minimal SU(5) model.  The decay rate can be written as:
\begin{eqnarray}
\Gamma(p \to \pi^{0} e^{+})& = & \alpha_{H}^{2} \frac{m_{p}}{64 \pi
  f_{\pi^{2}}}(1+D+F)^{2} \Bigl(\frac{g_{5}^{2}
  A_{R}}{M_{V}^{2}}\Bigr)^{2} \times \nonumber \\ 
& & (1+(1+|V_{ud}|^{2})^{2}) \label{eqn:minimaldimsix}
\end{eqnarray}
Here, $\alpha_{H}$, is the hadronic matrix element, evaluated to the
JLQCD collaboration to be $\alpha_{H}=0.015 \pm .001$ GeV$^{3}$.
$A_R$ is a overall renormalization factor that contains both a long
and short-distance piece \cite{IbanezMunoz}. $F$ and $D$ are chiral
Lagrangian parameters.  The piece $(1+|V_{ud}|^2)^2$ comes from the
operator ${\bf 10}_i^* {\bf 10}_i {\bf 10}_j^* {\bf 10}_j$, while the
piece $1$ comes from the operator ${\bf 10}_i^* {\bf 10}_i
\overline{\bf 5}_1^* \overline{\bf 5}_1$.  Our numerical evaluation
yields:
\begin{eqnarray}
\lefteqn{
\frac{1}{\Gamma(p \to \pi^{0} e^{+})} = } \nonumber \\
& &8 \times 10^{34} \mbox{ yrs.} \times \Bigl( \frac{0.015
  \mbox{GeV$^{3}$}}{ \alpha_{H}} \Bigr)^{2} \Bigl(
\frac{M_{V}}{10^{16} \mbox{GeV}} \Bigr)^{4}. 
\end{eqnarray}
In section \ref{sec:RGE}, we constrained the product: $(M_{V}^{2}
M_{\Sigma})^{1/3}$.  We now try to disentangle the product.  The case
$M_{V} \gg M_{\Sigma}$ is perfectly allowed, and conceivably, the mass
of $M_{V}$ might be as high as the Planck Scale, so the dimension six
decay might be completely out of reach.  On the other hand, $M_{V}$
cannot be arbitrarily small.  ${\cal W} \ni \frac{f}{3} \mbox{Tr}
\Sigma^{3}$, and we can write $M_{\Sigma}=\frac{f M_{V}}{2 \sqrt{2}
  g_5}$.  Imposing the constraint that the Higgs trilinear
self-coupling, $f$, should not blow up before the Planck scale,
reference \cite{Hisano} found $M_{V} > 0.56 M_{\Sigma}$. Taken with
Eqn.~(\ref{eqn:GUTLimit}), we find that $M_{V} > 1.4 \times 10^{16}$
GeV.  If $M_{V}$ is indeed close to this limit, it is conceivable that
dimension six proton decay might be accessible at a next-generation
nucleon decay experiment.

The above discussion of dimension six decays can be easily modified to
discuss the flipped-SU(5) model \cite{flipped}. In this model,
dimension five operators are absent. However, the dimension six
operators arising from the exchange of $X$ bosons are still present.  In
this model, the scale of the $X$ bosons is determined solely by the
unification of the SU(2) and SU(3) couplings (the ``exact''
unification of the three couplings must be viewed as something of an
accident).  In this case, the decay rate becomes:
\begin{equation}
  \Gamma(p \to \pi^{0} e^{+})= \alpha_{H}^{2} \frac{m_{p}}{64 \pi
    f_{\pi^{2}}}(1+D+F)^{2} \Bigl(\frac{g_{5}^{2}
    A_{R}}{M_{V}^{2}}\Bigr)^{2}.  
\end{equation}
This decay rate is is smaller than Eqn.~(\ref{eqn:minimaldimsix}) by
almost a factor of five, because only the ${\bf 10}_i^* {\bf 10}_i
\overline{\bf 5}_1^* \overline{\bf 5}_1$ operator contributes to this mode
and hence the factor of $(1+|V_{ud}|^{2})^{2}$ is absent.  (This point
had not been made in the literature to the best of our knowledge.)
However, it turns out that the mass of the gauge bosons, $M_{V}$, can
be lower than in the minimal SU(5) case, thereby allowing a higher decay rate for flipped-SU(5) theories.
Let us now determine how small $M_{V}$ can actually be.  In this case, we cannot
use the same method we used for minimal SU(5) to constrain the mass of
$M_{V}$, as the condition that only two couplings unify
is less stringent.  On the other had, there is no $\Sigma$ that gives
threshold corrections to the couplings.  So, by using the condition
that $\alpha_{2}$ and $\alpha_{3}$ unify, we can determine a bound
on the combination $(M_{V}^{2} M_{H_{C}})^{1/3}$.  We find
\begin{eqnarray}
\label{eqn:flippedbound}
3.3  \times 10^{15} \leq (M_{V}^{2} M_{H_C})^{1/3} \leq 8.2 \times 10^{15}
\mbox{ GeV}\nonumber \\ 
(90\% \mbox{ confidence level}).
\end{eqnarray}

Now, we expect that $M_{H_{C}}$ should be near (or below) the GUT scale, as it
arises from a coupling times a GUT scale vacuum expectation value.  Using this peturbativity argument, reference \cite{Hisano} has shown that $M_{H_{C}} < 2.0 \, M_{V}$.  Applying this result in Eqn.~(\ref{eqn:flippedbound}), we find that $M_{V} > 2.6 \times 10^{15}$ GeV.  On the other hand, the Super Kamiokande bound\cite{SKPhD} on the $p \to \pi^{0} e^{+}$ channel of $\tau_{p} > 2.6 \times 10^{33}$ years translates into a limit of $M_{V} > 2.8 \times 10^{15}$ GeV.  Therefore, current nucleon decay experiments have just begun to probe the dimension-six operators of the flipped-SU(5) model.

\section{Conclusions}

In conclusion, we find that by forcing the gauge couplings to unify,
we can place a rather stringent bound on the colored Higgs mass in the
Minimal SUSY SU(5).  A more precise determination of $\alpha_s(m_Z)$
has greatly improved this bound.  In light of this, LEP has done a
great deal to constrain a SUSY SU(5) theory.  Using the constraint on
the colored Higgs, we find that the minimal SUSY SU(5) grand unified
theory has been easily excluded by the Super Kamiokande experiment.
Even a scenario allowing for heavy scalars in the first two
generations does not allow SU(5) to avoid the experimental bounds.

However, we have also mentioned several theoretical approaches that
can substantially suppress the dimension five decay.  It is not yet
possible to exclude these options.  So, while it is is impossible to
say that no SU(5) theory is correct, it is correct to say the the
minimal SUSY SU(5) theory is excluded, even if the superpartners are
taken to be very heavy.  It is hoped that future nucleon decay
experiments can probe the dimension six operators in the future,
providing conclusive evidence for a grand unified theory.
   
\begin{acknowledgements}
  This work was supported in part by the Director, Office of Science,
  Office of High Energy and Nuclear Physics, Division of High Energy
  Physics of the U.S. Department of Energy under Contract
  DE-AC03-76SF00098 and in part by the National Science Foundation
  under grant PHY-95-14797.  AP is also supported by a National
  Science Foundation Graduate Fellowship.
\end{acknowledgements}


\begin{thebibliography}{99}
  
\bibitem{SKPDecay} Y.~Hayato {\it et al.}  [SuperKamiokande Collaboration],
  Phys.\ Rev.\ Lett.\ {\bf 83}, 1529 (1999) [hep-ex/9904020].
  
\bibitem{Cohen} A.~G.~Cohen, D.~B.~Kaplan and A.~E.~Nelson,
  Phys.\ Lett.\ B {\bf 388}, 588 (1996) [hep-ph/9607394].
  
\bibitem{Feng} J.~A.~Bagger, J.~L.~Feng, N.~Polonsky and R.~Zhang,
  Phys.\ Lett.\ B {\bf 473}, 264 (2000) [hep-ph/9911255].
  
\bibitem{HKN} J.~Hisano, K.~Kurosawa and Y.~Nomura,
  Nucl.\ Phys.\ B {\bf 584}, 3 (2000) [hep-ph/0002286].
  
\bibitem{Ellis} J.~R.~Ellis, D.~V.~Nanopoulos and S.~Rudaz,
  Nucl.\ Phys.\ B {\bf 202}, 43 (1982).
  
\bibitem{GotoNihei} T.~Goto and T.~Nihei,
  Phys.\ Rev.\ D {\bf 59}, 115009 (1999) [hep-ph/9808255].
  
\bibitem{Nath1} P.~Nath, A.~H.~Chamseddine and R.~Arnowitt,
  Phys.\ Rev.\ D {\bf 32}, 2348 (1985).
  
\bibitem{Nath2} P.~Nath and R.~Arnowitt,
  Phys.\ Rev.\ D {\bf 38}, 1479 (1988).
  
\bibitem{Hisano} J.~Hisano, H.~Murayama and T.~Yanagida,
  Nucl.\ Phys.\ B {\bf 402}, 46 (1993) [hep-ph/9207279].

\bibitem{WeinbergSakai} S.~Weinberg,
  Phys.\ Rev.\ D {\bf 26}, 287 (1982);\\
  N.~Sakai and T.~Yanagida,
  Nucl.\ Phys.\ B {\bf 197}, 533 (1982).
  
\bibitem{couplings} J.~Hisano, H.~Murayama and T.~Yanagida,
  Phys.\ Rev.\ Lett.\  {\bf 69}, 1014 (1992).

\bibitem{Hisano2} J.~Hisano, T.~Moroi, K.~Tobe and T.~Yanagida,
  Mod.\ Phys.\ Lett.\ A {\bf 10}, 2267 (1995) [hep-ph/9411298].

\bibitem{Einhorn} M.~B.~Einhorn and D.~R.~Jones,
  Nucl.\ Phys.\ B {\bf 196}, 475 (1982);\\
S.~P.~Martin and M.~T.~Vaughn,
Phys.\ Rev.\ D {\bf 50}, 2282 (1994)
[hep-ph/9311340].

\bibitem{Yamada} Y.~Yamada,
  Z.\ Phys.\ C {\bf 60}, 83 (1993).
  
\bibitem{LEPHiggs} LEPHWG, ALEPH, DELPHI, L3 and OPAL experiments,
  note LHWG/01-04 (preliminary), \url{http://lephiggs.web.cern.ch/LEPHIGGS/papers/July2001_MSSM/index.html}.
  
\bibitem{PDG} D.~E.~Groom {\it et al.}  [Particle Data Group Collaboration],
  Eur.\ Phys.\ J.\ C {\bf 15}, 1 (2000),
  available on the PDG WWW pages \url{http://pdg.lbl.gov/}.

\bibitem{DRThresh} I.~Antoniadis, C.~Kounnas and K.~Tamvakis,
  Phys.\ Lett.\ B {\bf 119}, 377 (1982).

\bibitem{DR} S.~P.~Martin and M.~T.~Vaughn,
  Phys.\ Lett.\ B {\bf 318}, 331 (1993) [hep-ph/9308222].

\bibitem{LEPChargino} LEPSUSYWG, ALEPH, DELPHI, L3 and OPAL
  experiments, note LEPSUSYWG/01-03.1 (preliminary), \url{http://lepsusy.web.cern.ch/lepsusy/www/inos_moriond01/charginos_pub.html}.
  
\bibitem{JLQCD} Y.~Kuramashi  [JLQCD Collaboration],
  talk at 2nd Workshop on Neutrino Oscillations and Their Origin (NOON
  2000), Tokyo, Japan, 2000, hep-ph/0103264;\\
  S.~Aoki {\it et al.}  [JLQCD Collaboration],
  Phys.\ Rev.\ D {\bf 62}, 014506 (2000) {\tt hep-lat/9911026}.
  
\bibitem{AM} N.~Arkani-Hamed and H.~Murayama,
  Phys.\ Rev.\ D {\bf 56}, 6733 (1997) [hep-ph/9703259].
  
\bibitem{ExtraDim} Y.~Kawamura,
  Prog.\ Theor.\ Phys.\ {\bf 105}, 999 (2001) [hep-ph/0012125];\\
  L.~Hall and Y.~Nomura,
  Phys.\ Rev.\ D {\bf 64}, 055003 (2001) [hep-ph/0103125];\\
  Y.~Nomura, D.~Smith and N.~Weiner,
  hep-ph/0104041.

\bibitem{BabuBarr} K.~S.~Babu and S.~M.~Barr,
  Phys.\ Rev.\ D {\bf 48}, 5354 (1993) [hep-ph/9306242].

\bibitem{flipped} S.~M.~Barr,
  Phys.\ Lett.\ B {\bf 112}, 219 (1982). \\
  J.~P.~Derendinger, J.~E.~Kim and D.~V.~Nanopoulos,
  Phys.\ Lett.\ B {\bf 139}, 170 (1984). \\
  I.~Antoniadis, J.~Ellis, J.~S.~Hagelin and D.~V.~Nanopoulos,
  Phys.\ Lett.\ B {\bf 194}, 231 (1987). \\
  J.~R.~Ellis, J.~L.~Lopez and D.~V.~Nanopoulos,
  Phys.\ Lett.\ B {\bf 371}, 65 (1996), {\tt hep-ph/9510246}.

\bibitem{PQsuppression} J.~Hisano, H.~Murayama and T.~Yanagida,
  Phys.\ Lett.\ B {\bf 291}, 263 (1992).
  
\bibitem{missingpartner} A.~Masiero, D.~V.~Nanopoulos, K.~Tamvakis and
  T.~Yanagida,
  Phys.\ Lett.\ B {\bf 115}, 380 (1982);\\
  B.~Grinstein,
  Nucl.\ Phys.\ B {\bf 206}, 387 (1982).

\bibitem{HagiwaraYamada} K.~Hagiwara and Y.~Yamada,
  Phys.\ Rev.\ Lett.\  {\bf 70}, 709 (1993).

\bibitem{Hisano3} J.~Hisano, T.~Moroi, K.~Tobe and T.~Yanagida,
  Phys.\ Lett.\ B {\bf 342}, 138 (1995) [hep-ph/9406417].
  
\bibitem{BPW} K.~S.~Babu, J.~C.~Pati and F.~Wilczek,
Nucl.\ Phys.\ B {\bf 566}, 33 (2000) [hep-ph/9812538].

\bibitem{HMN} L.~J.~Hall, H.~Murayama and Y.~Nomura,
  hep-th/0107245.
  
\bibitem{hisanorecent} J.~Hisano,
  {\tt hep-ph/0004266}, talk given at Workshop on Neutrino
  Oscillations and Their Origin, Fijiyoshida, Japan, 11-13 Feb 2000.
  
\bibitem{IbanezMunoz} L.~E.~Ibanez and C.~Munoz,
  Nucl.\ Phys.\ B {\bf 245}, 425 (1984).

\bibitem{SKPhD}
Brett M. Viren, PhD Thesis, State University of New York at Stony Brook, May 2000, \url{http://www-sk.icrr.u-tokyo.ac.jp/doc/sk/pub/index.html}.
\end{thebibliography}
\end{document}